\documentclass[twocolumn,showpacs,amsmath,amssymb,prb]{revtex4}
\usepackage{graphicx,amstext}% Include figure files
\usepackage{dcolumn}% Align table columns on decimal point
\usepackage{bm}% bold math
\newcommand{\vs}{{\it vs.\@}}

\newcommand{\al}{{\it et al.\@}}

\newcommand{\la}{La$_{3}$Sr$_{3}$Ca$_{8}$Cu$_{24}$O$_{41}$}
\newcommand{\bq}{\begin{equation}}
\newcommand{\eq}{\end{equation}}
\begin{document}
\title{Variable-range hopping conductivity in the copper-oxygen chains
of La$_{3}$Sr$_{3}$Ca$_{8}$Cu$_{24}$O$_{41}$}

\author{T. Vuleti\'{c}}
 \email{tvuletic@ifs.hr}
\author{B. Korin-Hamzi\'{c}}
\author{S. Tomi\'{c}}
 \homepage{http://www.ifs.hr/real_science}
\affiliation{Institut za fiziku, P.O. Box 304, HR-10001 Zagreb,
Croatia}
\author{B. Gorshunov}
 \altaffiliation {Permanent address: General Physics Institute, Russian Academy
of Sciences, Moscow, Russia.}
\author{P. Haas}
\author{M. Dressel}
\affiliation{1.\@ Physikalisches Institut, Universit\"{a}t
Stuttgart, D-70550 Stuttgart, Germany}
\author{J. Akimitsu}
\author{T. Sasaki}
\affiliation{Department of Physics, Aoyama-Gakuin University,
Setagaya-ku, Tokyo 157-8572,  Japan}
\author{T. Nagata}
\affiliation{Department of Physics, Ochanomizu University,
Bunkyo-ku, Tokyo 112-8610,  Japan}

\date{\today}

\begin{abstract}
We show that the spin chain/ladder compound \la~is an insulator
with hopping transport along the chains. In the temperature range
35 - 280 K, DC conductivity $\sigma_{DC}(T)$ follows Mott's law of
variable-range hopping conduction; the frequency dependence has
the form $\sigma(\nu, T) = \sigma_{DC}(T) + A(T)\cdot \nu^{s}$,
where $s\approx 1$. The conduction mechanism changes from
variable-range hopping to nearest-neighbor hopping around $T_{c}
=$~300 K. The chain array thus behaves like a one-dimensional
disordered system. Disorder is due to random structural
distortions of chains induced by irregular coordination of the
La/Sr/Ca ions.
\end{abstract}

\maketitle

One of the most outstanding properties of (La, Sr,
Ca)$_{14}$Cu$_{24}$O$_{41}$ family of quantum spin chain/ladder
compounds is the superconductivity (SC) established in
Sr$_{0.4}$Ca$_{13.6}$Cu$_{24}$O$_{41}$ at 12 K under
pressure~\cite{Uehara96}. The parent material of this cuprate
superconductor, Sr$_{14}$Cu$_{24}$O$_{41}$, is a charge density
wave (CDW) insulator with a spin
gap~\cite{Gorshunov02,Blumberg02,Kumagai97}. Substituting
isovalent Ca for Sr suppresses the CDW insulating
phase~\cite{Vuletic03}, while the spin gap remains
finite~\cite{Kumagai97}. The latter indicates that SC is driven by
the spin-liquid state, in accord with theoretical
expectations~\cite{Dagotto92}.

In addition to the two-leg ladders, responsible for the
conductivity and superconductivity, the system comprises
one-dimensional (1D) CuO$_{2}$ chains (c-direction) and the (La,
Sr, Ca) layers. The chains are charge-reservoir from which holes
are transferred into the ladders keeping the average copper
valence unchanged. For the fully doped compound
Sr$_{14}$Cu$_{24}$O$_{41}$, which contains six holes per formula
unit, approximately five holes are observed in the chain
subsystem. In this case, the antiferromagnetic dimer pattern is
created in chains together with the charge order, both inducing
gaps in the spin and charge sector,
respectively~\cite{Regnault99,Kataev01}. Spin dimers are formed
between those Cu$^{2+}$ spins that are separated by a localized
Zhang-Rice singlet (Cu$^{3+}$), that is, by a site occupied by a
localized hole.

No definite understanding has been reached yet on the nature of
the spin/charge state and in particular on the charge dynamics in
the chain subsystem for various doping levels. The
La$_{6-y}$Sr$_{y}$Ca$_{8}$Cu$_{24}$O$_{41}$ compound provides a
good opportunity for such studies since here for intermediate
doping levels all holes reside on the chain sites~\cite{Nucker00}
and no spin gap is observed~\cite{Carter96}. Both susceptibility
and DC resistivity, measured on polycrystalline samples, increase
with lowering temperature, while their absolute values show a
strong decrease on strontium-doping~\cite{Carter96}. In
particular, the absolute value of the susceptibility is directly
correlated with the number of Cu$^{2+}$ in chains, indicating that
doped holes reside entirely on chains.

In order to understand the nature of the spin/charge state in the
spin chains with intermediate hole doping, there is a need to
clarify the mechanism of the charge transport. In this
communication we present conductivity measurements on single
crystalline La$_{3}$Sr$_{3}$ Ca$_{8}$Cu$_{24}$O$_{41}$ in a wide
frequency and temperature range. We show that the conductivity
measured between 280 K and 35 K obeys a true variable-range
hopping law as in disordered non-crystalline insulators; in this
way we exclude the existence of the charge order pattern found in
the fully doped spin chains. In addition, we find no signature of
the CDW-related dielectric response; this fact represents definite
evidence that the CDW insulating phase observed in
Sr$_{14-x}$Ca$_{x}$Cu$_{24}$O$_{41}$ is established in the ladder
sub-unit~\cite{Gorshunov02,Vuletic03}.

DC resistivity was measured between 35 K and 700 K. In the
frequency range $\nu = 0.1$~Hz -- 1 MHz the complex conductance
was measured; for $\nu < 100$~Hz we used a set-up for
high-impedance samples while for 20 Hz~$<\nu<$~1 MHz a Hewlett
Packard HP4284A impedance analyzer was utilized~\cite{Pinteric01}.
The data at the lowest frequency match our four-probe DC
measurements. At frequencies $\nu=6$~-- 10000 cm$^{-1}$ the
complex dielectric function was obtained by a Kramers-Kronig
analysis of the reflectivity and by complex transmission
measurements~\cite{Kozlov98} at the lowest frequencies 6 -- 20
cm$^{-1}$. All measurements were done along the crystallographic
c-axis of high-quality single crystal.

%   FIGURE 1
\begin{figure}
\centering\includegraphics[clip,scale=0.36]{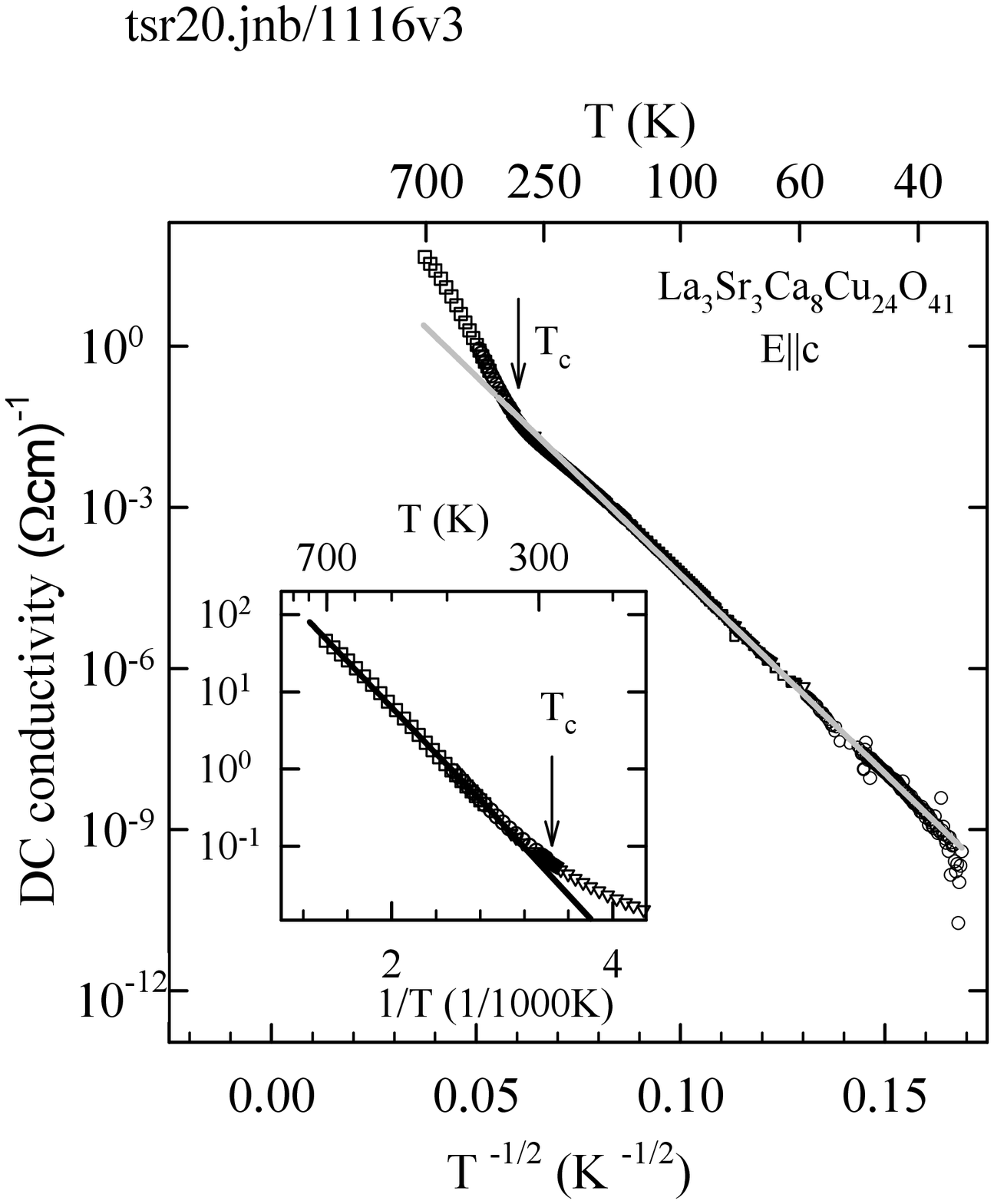} \caption{DC
conductivity  $\sigma_{DC}$ of \la~single crystal along the
crystallographic c direction \vs~$T^{-1/2}$.  $\sigma_{DC}$
follows a simple activation behavior above $T_c = 300$~K (full
line in the inset) indicating the nearest-neighbor hopping. Below
$T_c$ the behavior  $\sigma_{DC}\propto \exp(T^{-1/2})$ is
observed (full line) corresponding to the regime of variable-range
hopping in one dimension. $T_c$ is determined by the
crossing of extrapolated fitting curves, with the error bar of $\pm15$~K.} \label{Fig1}
\end{figure}

 Fig.~\ref{Fig1} shows the behavior of DC conductivity in the wide temperature
range from 35 K (the lowest temperature obtained in our
experiment) up to 700 K. Above $T_{c} = 300$~K, the DC
conductivity follows a simple activation behavior $\sigma_{DC}(T)
\approx \exp (-2\Delta/T)$ with $2\Delta = 3200$~K (see inset). As
directly seen from the $\log \sigma_{DC}(T)$ \vs~$T^{-1/2}$ plot,
presented in the main panel, below $T_{c}$ down to 35 K the
conductivity perfectly follows the variable-range hopping (VRH)
behavior
\bq 
\sigma_{DC}(T) = \sigma_{0}\exp{\left[ - \left({T_0/T}\right)^{\frac{1}{1+d}}\right]} 
\label{Eq1}
\eq 
with the dimensionality of the system $d = 1$. These results
clearly demonstrate the hopping mechanism of charge transport in
one dimension; at $T_{c} =300$~K it crosses over from
nearest-neighbor hopping to variable-range hopping. The cross-over
temperature $T_{c}$ is given by $T_{c} = \Delta/(2\alpha c)$
~\cite{MottDavis,Yu01}. Here the energy of sites near the Fermi
energy available for hops, has an uniform distribution in the
range  $-\Delta$ to $+\Delta$, $c$ is the distance between the
nearest Cu chain sites and $\alpha^{-1}$ is the localization
length. By using $\Delta = 1600$~K and $c = 2.77 {\AA}$, we find
$\alpha^{-1} \approx 1 {\AA}$. Finally, the value of the VRH
activation energy $T_{0}^{exp} = 2.9 \cdot 10^{4}$~K, obtained
from the fit of our data to Eq.~\ref{Eq1}, is very close to the
one expected theoretically: $T_{0}^{th} = 8 \Delta c \alpha
\approx 3.5 \cdot 10^{4}$~K.

%   FIGURE 2
\begin{figure}
\centering\includegraphics[clip,scale=0.37]{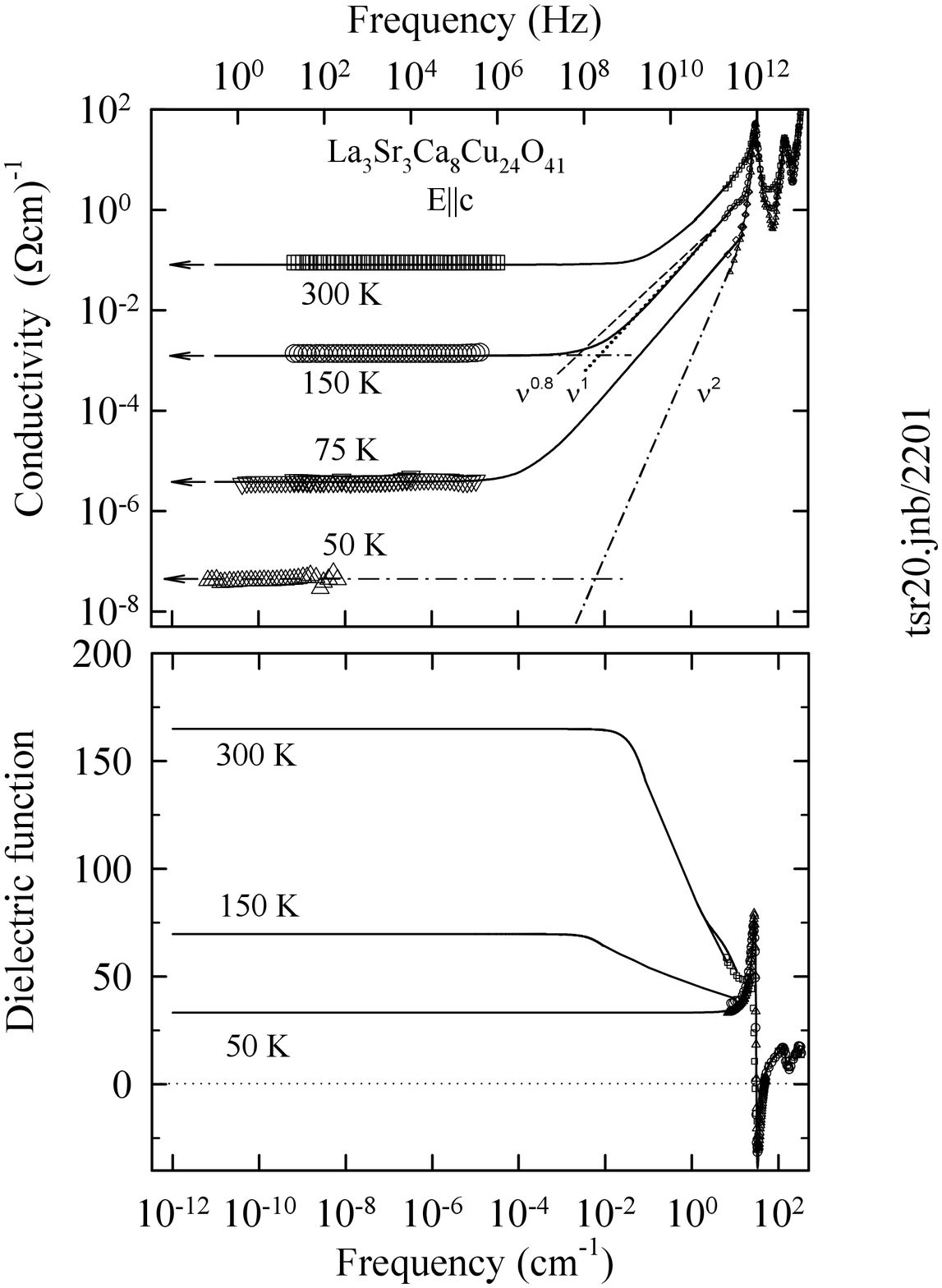} \caption{Upper
panel: broad-band conductivity spectra of \la~single crystal along
the c axis for four representative temperatures. Hopping
conduction of the form  $\nu^s$, $s \approx 1$, is found between 6
and 20 cm$^{-1}$ for $T\geq75$~K, while a frequency independent
behavior is found in the radio-frequency range for all
temperatures. The arrows denote the DC values. The full lines are
fits to the form  $\sigma(\nu,T) = \sigma_{DC}(T) + A(T)\nu^1$. A
pure power law contributions with $s = 0.8$ and 1 are shown for $T
= 150$~K by a dashed and dotted lines, respectively. At 50 K only
a $\nu^2$ contribution (dashed-dotted line) due to the low-energy
phonon tail is observed. Lower panel: dieletric function of
\la~single crystals along the c axis (open points). The lines
correspond to the fit of the conductivity spectra shown in the
upper panel as described in the text. At $T = 50$~K, only phonon
contribution to dielectric function is found.} \label{Fig2}
\end{figure}

Fig.~\ref{Fig2} demonstrates the conductivity $\sigma (\nu, T)$
and the dielectric function $\varepsilon'(\nu,T)$ spectra of
La$_{3}$Sr$_{3}$Ca$_{8}$Cu$_{24}$O$_{41}$ over the entire
frequency range for different temperatures. We present here only
the spectra up to 350 cm$^{-1}$ including a couple of phonon
lines. As confirmed by our fit (see below), the kinks in $\sigma
(\nu, T)$ and $\varepsilon'(\nu,T)$ on the left side of the
lowest-frequency phonon are of electronic (non-phonon) origin and
we assign this contribution to the hopping of holes in the chains.
Excluding the phonon component, the electronic conductivity can be
expressed as the sum of two terms \bq \sigma(\nu,T) =
\sigma_{DC}(T)+A(T)\cdot \nu^s \qquad s\approx1 \label{Eq2} \eq
where $\sigma_{DC}(T)$ is given by Eq.~\ref{Eq1}. We note that the
frequency independent behavior is found in the radio-frequency
range for all temperatures (open symbols in Fig.~\ref{Fig2}).
Similar dependences have been observed in a variety of disordered
systems~\cite{Dyre00}.  The frequency-dependent component
$\sigma_{AC}(\nu,T) = A(T) \cdot \nu^{s}$ is found to contain a
temperature dependent prefactor $A(T)$. The cross-over frequency
$\nu_{co}$ from the frequency independent to the frequency
dependent conductivity can be estimated from the condition that
the AC hopping length has to be smaller than the DC hopping length
in order that $\sigma_{AC}(\nu,T)$ overcomes
$\sigma_{DC}(T)$~\cite{MottDavis}. For one-dimensional VRH, the DC
hopping length is given by $R_{0} = (\Delta c/2\alpha T)^{1/2}$,
and the AC hopping length is $R_{\nu} = \frac{1}{2}\alpha \ln
(\nu_{ph} / \nu_{co})$, where the attempt frequency $\nu_{ph}$
depends on the electron-phonon interaction. Assuming
$\nu_{ph}\approx 10^{12}$ s$^{-1}$, we find for the cross-over
frequency $\nu_{co}$ the values 0.15 cm$^{-1}$, 0.015 cm$^{-1}$
and 0.0006 cm$^{-1}$ for $T =$~300 K, 150 K and 75 K,
respectively. These values coincide nicely with those obtained
when the $\nu^{s}$ fits in 6 -- 20 cm$^{-1}$ range are
extrapolated to lower frequencies (Fig.~\ref{Fig2}). In
particular, the choice of the exponent $s = 1$ appears to be the
most appropriate. For example, at $T = 150$~K we find
$\nu_{co}\approx 0.0027$ cm$^{-1}$ and 0.013 cm$^{-1}$, for $s =
0.8$ and 1, respectively. Experimentally we are not able to
distinguish between the $\nu^{0.8}$ and $\nu^{1}$ dependences due
to relatively narrow frequency range in which $\sigma \sim \nu^s$
behavior is detected. At $T \leq 50$~K the hopping vanishes
because the charge carriers are frozen out, and we observe only
the $\nu^{2}$ contribution to the conductivity associated with the
low-energy phonon (Lorentzian) tail.

Next, we estimate the dielectric constant $\varepsilon$'
associated with the hopping conduction. In order to do so we fit
the conductivity spectra using the Drude term for the frequency
independent plateaus below 10$^{6}$ Hz and a set of Lorenztians to
smoothly describe the increase of $\sigma_{AC}(\nu)$ at higher
frequencies. The results are shown by solid lines in
Fig.~\ref{Fig2}; the low-frequency dielectric constant is about
170 at 300 K and decreases with lowering the temperature. We were
not able to extract the $\varepsilon$' values from our
radio-frequency measurements since the accuracy in determination
of $\varepsilon$' is $\pm300$ for the geometry of the samples
used.

Finally, we want to point out that no signature of the CDW-related
features of dielectric response in the radio-frequency range is
found. Since in this system all holes reside on chains, this
result yields definite evidence that the CDW insulating phase
observed in Sr$_{14-x}$Ca$_{x}$Cu$_{24}$O$_{41}$ is established in
the ladder sub-unit~\cite{Blumberg02}.

%   FIGURE 3
\begin{figure}
\centering\includegraphics[clip,scale=0.35]{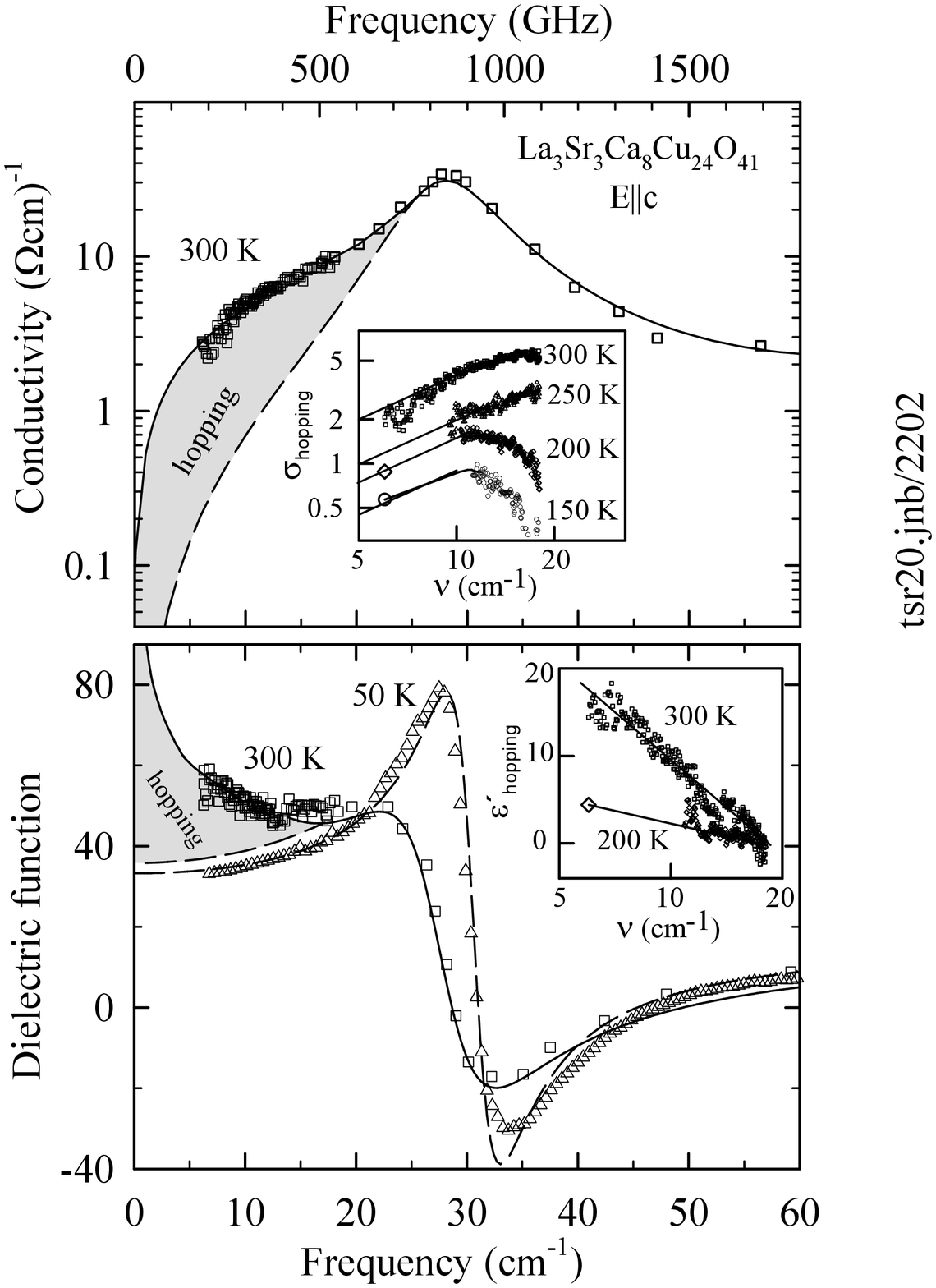} \caption{Optical
conductivity (upper panel) and dielectric function (lower panel)
of \la~single crystal measured along the c axis in the range 6 -
60 cm$^{-1}$. The dashed lines are the fits including only the
contributions associated to phonon at 28 cm$^{-1}$. The full lines
correspond to the fit which in addition includes an electronic
hopping conduction (see text). The insets show (for a few selected
temperatures) the conductivity and the dielectric function due to
hopping (phonons subtracted) which die out at low temperatures and
high frequencies. Full lines are fits to $A(T)\cdot \nu^1$. At T =
50 K, only phonon contribution is detected.} \label{Fig3}
\end{figure}

Fig.~\ref{Fig3} shows a blow-up of the optical conductivity and
dielectric function at frequencies 6 -- 60 cm$^{-1}$ where a
crossover from hopping to phonon-related response is observed. A
contribution due to hopping (shaded) is clearly identified in
addition to the phonon part (dashed line). The insets show the
difference between the total and the phonon-associated components
to the spectra of $\sigma$ and ${\varepsilon}$', which basically
is the pure hopping contribution. The expected dieing out of the
hopping contribution becomes more pronounced with lowering the
temperature and with increasing the frequency. The latter is - to
the best of our knowledge - observed for the first time.

We now comment on the hopping transport found in the chains of
La$_{3}$Sr$_{3}$Ca$_{8}$Cu$_{24}$O$_{41}$ in comparison with
disordered non-crystalline insulators. First, for the latter
compounds the value of the exponent in Eq.~\ref{Eq1} is commonly
found to be 1/4, corresponding to hopping in 3 dimensions; it
becomes 1/2 if electron-electron interaction plays a
role~\cite{Ladieu96}. However, electron-electron interactions are
expected to have a significant impact on the correlated
many-electron hopping only when the temperature T is larger than
$T_{0}$, which is far from experimental range since $T_{0} = 2.9
\cdot 10^{4}$~K, \cite{MottDavis,Knotek74}. The situation is different for systems consisting
of parallel chains of finite length where small disorder leads to weakly
localized states~\cite{Shante73}. This model gives the
temperature exponent 1/2 as found in our experiment. We conclude
that the exponent 1/2 confirms the one-dimensional nature of the
electronic structure of chains in
La$_{3}$Sr$_{3}$Ca$_{8}$Cu$_{24}$O$_{41}$, which is in accord with
the crystallographic structure~\cite{Siegrist88}. Second, the
obtained value for $T_{0}$ indicates the standard regime of the
VRH where the hopping distance $R_{0}$ is larger than the
localization length $\alpha^{-1}$; the extremely small
$\alpha^{-1} \approx 1 {\AA}$ shows that the system is far from
the metal-insulator transition. Following the usual interpretation
of the VRH law, from $T_{0} = 16 \alpha^{3}/ n(E_{F})$ we find the
electronic density of states at the Fermi level $n(E_{F})\approx
5.5 \cdot 10^{24}$~ eV$^{-1}$cm$^{-3}$. Finally, a straightforward
consequence of the observed VRH conduction is that its
extrapolation indicates a zero conductivity at $T=0$, in accord
with the theory developed for disordered non-crystalline
insulators.

Generally, a frequency-dependent conductivity varying as
$A(T)\cdot \nu^{s}$, where $s \approx 1$, does not necessarily
imply hopping conduction~\cite{MottDavis}. However, we suggest
that DC VRH conduction as well as the power-law AC conduction are
attributed to the same set of localized states near the Fermi
level. The value of the exponent close to one indicates that the
observed AC conductivity is due to phonon-assisted hops between
spatially distinct sites similarly to the DC contribution, and not
to the photon absorption for which $s \approx 2$ is usually
found~\cite{MottDavis}. Note that we did not find the latter in
the whole measured frequency range, despite the theoretical
prediction that with increasing frequency there is a cross-over
from the regime dominated by phonon-assisted
$\sigma_{hopping}\approx \nu^{s}$, $s \leq 1$ to that dominated by
photon-assisted conduction which varies as $\nu^{s}$, $s\approx
2$. The conductivity observed at 50 K, which follows a $\nu^{2}$
behavior, is in our case simply a phonon tail, and is not due to
hopping. Further, the observation of a strongly T-dependent
$\sigma_{hopping}(10 \mathrm{cm}^{-1})/ \sigma_{DC}$ ratio
confirms the idea that the hopping transport involves localized
states near the Fermi energy. In addition, we find that the
prefactor $A(T)$ follows the linear $T$ dependence for $T \geq
60$~K. This indicates that the thermal energy $k_{B}T$ is small
compared to the energy range over which $n(E_{F})$ may be taken as
the constant~\cite{MottDavis}, in agreement with the estimates for
the bandwidth associated with chains~\cite{Arai97}. However, in
this case $A(T) = 0$ for $T<60$~K which implies no phonon-assisted
contribution to $\sigma(\nu)$ at low temperatures. This seems
surprising since we observed the VRH law for the DC conductivity
down to at least 35 K. The other possibility is that $A(T)$
follows a $T^{1.8}$ law at all temperatures. We note that the
exponent $s = 1.8$ is larger than $0 \leq s \leq 1$
expected~\cite{MottDavis} for conventional disordered
non-crystalline insulators. Finally, we want to comment on a
decrease of the phonon-assisted hopping conductivity observed at
high frequencies 6 -- 20 cm$^{-1}$ (insets of Fig.~\ref{Fig3}). In
the theoretical two-site hopping model, at frequencies of the
order of $\nu_{ph}\approx 10^{12}$ s$^{-1}$, the frequency
dependence of $\sigma_{hopping}$ saturates due to the fact that
the exponent $s$ decreases logarithmically with increasing
frequency~\cite{Bottger85}. Our observation is in line with this
theoretical prediction. Moreover, the final fading out of the
phonon-assisted hopping at very high frequencies is intuitively
expected and deserves more theoretical attention.

 At microscopic scale, we propose that strong local distortions
of the chains due to irregular coordination of La$^{3+}$,
Sr$^{2+}$ and Ca$^{2+}$ ions~\cite{Siegrist88} induce a
non-periodic potential in which holes reside. The finding of the VRH law in the
measured conductivity can be then viewed as a result of distorted distribution
of microscopic conductivities, as predicted in Anderson localization theories.
Therefore, copper-oxygen chains in partially doped
La$_{3}$Sr$_{3}$Ca$_{8}$Cu$_{24}$O$_{41}$ can be considered as a
system in which disorder, associated with random distribution of
holes, causes the Anderson localization. This is in contrast to
the observations for chain sub-unit in fully doped
Sr$_{14}$Cu$_{24}$O$_{41}$, where a charge gap opens due to the
charge order developed in conjunction with the antiferromagnetic
dimer pattern. We propose that the copper-oxygen chain sub-unit
behaves like a one-dimensional disorder-driven insulator for the
whole range of intermediate hole counts $0<n_{h}<6$ ~\cite{La52}, and crosses
over into a charge-ordered gapped insulator at full doping $n_{h}
= 6$. Note that the latter phase is established in chains concomitantly with the CDW
gapped state in ladders. Moreover, both phases are
suppressed by calcium-doping at seemingly similar rates,
indicating a profound interplay between chain and ladder
sub-units~\cite{Vuletic03,Kataev01}. 

 In conclusion, the investigations of the frequency and temperature
dependent conductivity yield clear evidence for variable-range
hopping transport in chains of a spin-chain/ladder system
La$_{3}$Sr$_{3}$Ca$_{8}$Cu$_{24}$O$_{41}$. The absence of holes in
ladders for intermediate hole counts eliminates the CDW phase in
ladders, and suppresses the charge-ordered gapped state in chains
in favor of disorder-driven insulating phase. These results reveal an intriguing
possibility for the existence of a phase transition close to
$n_h=6$ in the phase diagram of (La,Sr,Ca)$_{14}$Cu$_{24}$O$_{41}$ compounds.
Further experiments on materials with very low La content, which corresponds to
$n_h\lesssim6$ should elucidate our proposal.

We thank G. Untereiner for the samples preparation. This work was
supported by the Croatian Ministry of Science and Technology and
the Deutsche Forschungsgemeinschaft (DFG).


\begin{thebibliography}{10}
%1
\bibitem{Uehara96}
M. Uehara \al,
\newblock{\em J.Phys.Soc.Japan} {\bf65}, 2764 (1996).
%2
\bibitem{Gorshunov02}
B. Gorshunov \al,
\newblock{\em Phys.Rev.B} {\bf66}, 060508(R) (2002).
%2,5
\bibitem{Blumberg02}
G. Blumberg \al,
\newblock { Science} {\bf297}, 584 (2002).
%3
\bibitem{Kumagai97}
K. Kumagai \al,
\newblock{\em Phys.Rev.Lett.} {\bf78}, 1992 (1997).
%4
\bibitem{Vuletic03}
T.Vuleti\'{c} \al,
\newblock{submitted to {\em Physical Review Letters}}
%5
\bibitem{Dagotto92}
E.Dagotto \al,
\newblock{\em Phys.Rev.B} {\bf45}, 5744 (1992).
%6
\bibitem{Regnault99}
L.P.Regnault \al,
\newblock{\em Phys.Rev.B} {\bf59}, 1055 (1999).
%7
\bibitem{Kataev01}
V.Kataev \al,
\newblock{\em Phys.Rev.B} {\bf64}, 104422 (2001).
%8
\bibitem{Nucker00}
N. N\"{u}cker \al,
\newblock{\em Phys.Rev.B} {\bf62}, 14384 (2000).
%9
\bibitem{Carter96}
S.A.Carter \al,
\newblock{\em Phys.Rev.Lett.} {\bf77}, 1378 (1996).
%10
\bibitem{Pinteric01}
M. Pinteri\'{c} \al,
\newblock{\em Eur. Phys. J. B} {\bf 22}, 335 (2001).
%11
\bibitem{Kozlov98}
G. Kozlov and A. Volkov, {\em Topics in Applied Physics}, Vol. 74,
Millimeter and Submillimeter Wave Spectroscopy of Solids, ed.~G.Gr\"{u}ner (Springer-Verlag, Berlin, 1998).
%13
\bibitem{MottDavis}
N.F.Mott and E.A.Davis, {\em Electronic Processes in
Non-crystalline Solids} (Oxford University, London, 1971).
%14
\bibitem{Yu01}
Z.G.Yu and X.Song,
\newblock{\em Phys.Rev.Lett.} {\bf86}, 6018 (2001).
%15
\bibitem{Dyre00}
J.C.Dyre \al,
\newblock{\em Rev.Modern Physics} {\bf72}, 873 (2000).
%16
\bibitem{Ladieu96}
F.Ladieu \al,
\newblock{\em Ann.Phys. (Paris)} {\bf21}, 267 (1996).
%17
\bibitem{Knotek74}
M.L.Knotek and M.Pollack,
\newblock{\em Phys.Rev.B} {\bf9}, 664 (1974).
%18
\bibitem{Shante73}
V.K.S.Shante \al,
\newblock{\em Phys.Rev.B} {\bf8}, 4885 (1973).
%19
\bibitem{Siegrist88}
T.Siegrist \al,
\newblock{\em Mat.Res.Bull.} {\bf23}, 1429 (1988).
%20
\bibitem{Arai97}
M.Arai and H.Tsunetsugu,
\newblock{\em Phys.Rev.B} {\bf56}, R4305 (1997).
%21
\bibitem{Bottger85}
H.B\"{o}ttger and V.V.Bryskin, {\em Hopping Conduction in Solids}
(Akademie-Verlag, Berlin, 1985).
%21
\bibitem{La52}
We have observed the VRH transport with almost unchanged parameters ($T_c$ and
activation energy) also in La$_{5.2}$Sr$_{0.8}$Ca$_{8}$Cu$_{24}$O$_{41}$ with
hole count $n_h = 1$ on chains."

\end{thebibliography}
\end{document}